\documentclass[12pt]{iopart}

\newcommand{\del}{\partial}

\usepackage{graphicx}

\begin{document}

\title[Charge fluctuations in chiral models 
and the QCD  phase transition ]{ Charge fluctuations in chiral models and the QCD phase transition }

\author{Vladimir  Skokov}
\address  {  GSI Helmholtzzentrum f\"ur Schwerionenforschung, D-64291
Darmstadt, Germany  }
\ead{V.Skokov@gsi.de}

\author{Bengt  Friman}
\address  {  GSI Helmholtzzentrum f\"ur Schwerionenforschung, D-64291
Darmstadt, Germany  }

\author{Frithjof Karsch}
\address  {  Physics Department, Brookhaven National Laboratory, Upton, NY 11973, USA
\\
Fakult\"at f\"ur Physik, Universit\"at Bielefeld,
 D-33501 Bielefeld, Germany}

\author{Krzysztof Redlich}
\address  {   Institute of Theoretical Physics, University of Wroclaw, PL--50204
  Wroc\l aw, Poland 
  \\ ExtreMe Matter Institute EMMI, GSI, D-64291 Darmstadt, Germany}

\begin{abstract}
 We consider the Polyakov loop-extended two flavor
chiral quark--meson model and discuss critical phenomena related with the spontaneous breaking
of the chiral symmetry. The model is explored beyond the mean-field approximation in the framework of the functional renormalisation group. We  discuss properties of the net-quark number density
fluctuations as well as their higher cumulants. We  show that with the increasing net-quark number density,
the higher order cumulants exhibit  a strong sensitivity to the chiral crossover transition.
We discuss their role as probes of  the chiral phase transition in
heavy-ion collisions at RHIC and LHC.
\end{abstract}

\section{Introduction}

Lattice Quantum Chromodynamics (LQCD) \cite{L_lgt3}  { has}
confirmed that at finite temperature and  small baryon density  QCD { exhibits  restoration} of chiral symmetry and deconfinement.
However,  owing to the  fermion sign problem,
the thermodynamics of strongly interacting matter at large baryon densities  is { presently} not accessible { by first} principle
LQCD calculations. { A} viable framework for exploratory studies  of QCD  at finite density is however offered by
phenomenological models and effective theories.

The Polyakov loop extended Nambu--Jona--Lasinio (PNJL)   { and quark}--meson (PQM) { models}
reproduce essential features { of QCD} thermodynamics already in the mean-field approximation.
However, to correctly account for the critical behavior and scaling properties near the chiral phase transition,
{ thermal and quantum fluctuations must be included} in a non-perturbative manner.
This can be achieved e.g. by using methods based on the functional renormalization group (FRG)~\cite{Berges:review}.

We { compute} the cumulants of the net-quark number density ($\chi_n^B$) and electric charge density ($\chi_n^Q$)
at finite temperature and baryon chemical potential. We show that the higher-order cumulants exhibit a
{ distinctive structure in the vicinity of the transition, where they become negative. Such a structure follows from the scaling functions} of the  three dimensional O(4) universality class ~\cite{Friman:2011pf,karsch,Karsch:TP}.
Ratios of such { cumulants} have been suggested as sensitive probes for { the} critical behaviour and the rapid change of degrees of freedom at the
QCD  phase transition~\cite{Ejiri:2005wq}.

\section{Polyakov loop-extended quark--meson model  }

The quark--meson model is an effective realization of the low--energy
sector of QCD, which incorporates chiral symmetry. { By introducing the coupling} of  quarks
to a uniform temporal color gauge field represented by the Polyakov loop, the model { can be employed to explore}  properties  related { to color} confinement. 

In order to account for non-perturbative mesonic  fluctuations  in the PQM
model, we employ methods  based on the functional renormalization group (FRG).
The FRG  involves an infrared regularization of the fluctuations at a sliding momentum scale
$k$, resulting in a scale dependent
effective action $\Gamma_k$, the so-called effective average action~\cite{Berges:review}.
We treat the Polyakov loop
as a background field, which is introduced self-consistently on the
mean-field level { while the fluctuations of the quark and meson fields }are accounted for by solving the FRG flow equations.

  { The}
  flow equation for the scale-dependent grand canonical potential density, $\Omega_{k}=T\Gamma_{k}/V$, for { the
   quark and meson subsystem reads~\cite{Skokov:2010wb,Skokov:2010uh}}
  \begin{eqnarray}\label{eq:frg_flow}
    \del_k \Omega_k(\ell, \ell^*; T,\mu)&=&\frac{k^4}{12\pi^2}
    \left\{ \frac{3}{E_\pi} \Bigg[ 1+2n_B(E_\pi;T)\Bigg]
      +\frac{1}{E_\sigma} \Bigg[ 1+2n_B(E_\sigma;T)
      \Bigg]   \right. \\ \nonumber && \left. -\frac{4 N_c N_f}{E_q} \Bigg[ 1-
      N(\ell,\ell^*;T,\mu)-\bar{N}(\ell,\ell^*;T,\mu)\Bigg] \right\}.
  \end{eqnarray}
  Here $n_B(E_{\pi,\sigma};T)$ is the bosonic distribution function
  \begin{eqnarray*}
    &&n_B(E_{\pi,\sigma};T)=\frac{1}{\exp({E_{\pi,\sigma}/T})-1}, \\&&E_\pi = \sqrt{k^2+\overline{\Omega}^{\,\prime}_k}\;~,~ E_\sigma
    =\sqrt{k^2+\overline{\Omega}^{\,\prime}_k+2\rho\,\overline{\Omega}^{\,
        \prime\prime} _k};
  \end{eqnarray*}
where the primes denote derivatives with respect to $\rho = (\sigma^2+\vec{\pi}^2)/2$ of
  $\overline{\Omega}=\Omega+c\sigma$.
  The fermion distribution functions $N(\ell,\ell^*;T,\mu)$ and
  $ \bar{N}(\ell,\ell^*;T,\mu)=N(\ell^*,\ell;T,-\mu) $,
  \begin{eqnarray}\label{n1}
    N(\ell,\ell^*;T,\mu)&=&\frac{1+2\ell^*e^{\beta(E_q-\mu)}+\ell e^{2\beta(E_q-\mu)}}{1+3\ell e^{2\beta(E_q-\mu)}+
      3\ell^*e^{\beta(E_q-\mu)}+e^{3\beta(E_q-\mu)}}
    \label{n2}
  \end{eqnarray}
  are modified because of the  coupling to the gluon field. Finally, the quark energy is given by
    $E_q =\sqrt{k^2+2g^2\rho}$.

The minimum of the thermodynamic potential is determined by the
stationarity condition
\begin{equation}
  \left. \frac{d \Omega_k}{ d \sigma} \right|_{\sigma=\sigma_k}=\left. \frac{d
      \overline{\Omega}_k}{ d \sigma} \right|_{\sigma=\sigma_k} - c =0.
  \label{eom_sigma}
\end{equation}
The flow equation~(\ref{eq:frg_flow}) is solved numerically with the
initial cutoff $\Lambda=1.2$ GeV (see details in
Refs.~\cite{Skokov:2010wb,Skokov:2010uh}).  The initial conditions for the flow are
chosen to reproduce { { vacuum} properties:} the physical pion mass $m_{\pi}=138$
MeV, the pion decay constant $f_{\pi}=93$ MeV, the sigma mass
$m_{\sigma}=600$ MeV and the constituent quark mass $m_q=300$ MeV at
the scale $k\to 0$.  The symmetry breaking term, $c=m_\pi^2 f_\pi$,
corresponds to an external field { and does} not flow.

By solving Eq.~(\ref{eq:frg_flow}), one obtains the thermodynamic
potential for { the quark and meson  subsystem}, $\Omega_{k\to0} (\ell,
\ell^*;T, \mu)$, as a function of the Polyakov loop variables $\ell$
and $\ell^*$. The full thermodynamic potential $\Omega(\ell, \ell^*;T,
\mu)$ in the PQM model, including quark, meson, and gluon
degrees of freedom  is obtained by adding the effective gluon potential ${\cal U}(\ell,
\ell^*)$  to $\Omega_{k\to0} (\ell, \ell^*;T, \mu)$:
\begin{equation}
  \Omega(\ell, \ell^*;T, \mu) = \Omega_{k\to0} (\ell, \ell^*;T, \mu) + {\cal U}(\ell, \ell^*).
  \label{omega_final}
\end{equation}
At a given temperature and chemical potential, the Polyakov loop
variables, $\ell$ and $\ell^*$, are then determined by the stationarity
conditions:
\begin{equation}
  \label{eom_for_PL_l}
  \frac{ \partial   }{\partial \ell} \Omega(\ell, \ell^*;T, \mu)  =0, \quad  \frac{ \partial   }{\partial \ell^*}  \Omega(\ell, \ell^*;T, \mu)   =0.
  \label{eom_for_PL_ls}
\end{equation}

\section{Cumulants of conserved charge }

The fluctuations of conserved charges are quantified  by   cumulants $\chi_n^{B,Q}$,
which are  generalized susceptibilities { obtained by taking  derivatives of the} pressure $p=T^4 \hat{p}$
with respect to the corresponding chemical potential $\mu_{B,Q} = T \hat{\mu}_{B,Q} $:  $\chi_n^{B,Q} = \frac{ \partial^n \hat{p} }  {\partial \hat{\mu}_{B,Q}^n}$.
 
The cumulants of conserved charges are sensitive probes of the chiral phase
transition. They indicate the position, the order,  and in case of the
second-order phase transition the universality class of the corresponding phase transition.
The net baryon number density $n_B$ is discontinuous at a first-order
transition, whereas the susceptibility $\chi_2^B$ and higher cumulants
diverge at the critical end point~\cite{Stephanov:2008qz} {  and at the spinodal lines  of a first-order chiral phase transition~\cite{CS}.
In the chiral limit and at non-zero chemical potential, all generalized susceptibilities
$\chi_n^B$ with $n>2$ diverge at the $O(4)$ chiral critical
line~\cite{Ejiri:2005wq,Stephanov:2011pb}, while at vanishing  chemical potential
this divergence shows up only for  $n>4$.}

Close to the phase transition the sixth and eighth order cumulants decrease from { their
positive} values { in} the hadron resonance gas to negative values, in accordance with the 
expected dominance of { the O(4) singularity}~\cite{Friman:2011pf}. Since { the  fourth order cumulant
is unaffected by chiral critical phenomena at zero chemical potential, it remains positive for small chemical potentials}, as in the hadron resonance gas (HRG).
{ However, at} larger densities, { it picks up} a non-trivial contribution { from 
higher} order { $\mu=0$ cumulants, as seen in} the Taylor expansion $\chi^B_4(\hat{\mu}) =  \chi_4^B(0) + \frac12 \chi_6^B(0) \hat{\mu}^2 +O(\hat{\mu}^4) $.
Thus, the fourth order cumulant and { consequently} the kurtosis ($\chi_4^B/\chi_2^B$)  may become negative
close to the crossover  transition even if the critical { end} point does not exist \cite{Friman:2011pf,Stephanov:2011pb,Stephanov:TP}.
The { location of the} first zeros of $\chi_n^B$ for $n=4,6$ and 8  are shown in Fig.~1.
{ The middle panel shows that the} region, { where} the sixth order cumulant of { the} net baryon number fluctuations
is negative, is  { closely} correlated  with the crossover transition.

The FRG method  can { also be applied   to  compute the fluctuations  of electric charge} in the PQM model. 
Here the differences { relative to} the mean field calculations are essential, since  the FRG approach  
 { accounts for}
charged pion contributions to  $\chi^Q_n$. { In Fig.~1  we show} the range  of negative  $\chi^Q_6$, which
is  similar to that { found for}  $\chi^B_6$.  Thus, negative fluctuations of the sixth order moments of net baryon  as well as  net electric charge fluctuations can { both be} attributed { to  chiral critical dynamics}.
Consequently, { the experimental observation of negative sixth} order cumulants in heavy ion collisions at RHIC and LHC { would}  indicate  that the chemical freeze-out { takes place} in the vicinity of the chiral crossover transition \cite{Friman:2011pf}.

\begin{figure*}[t]
  \includegraphics*[width=5cm]{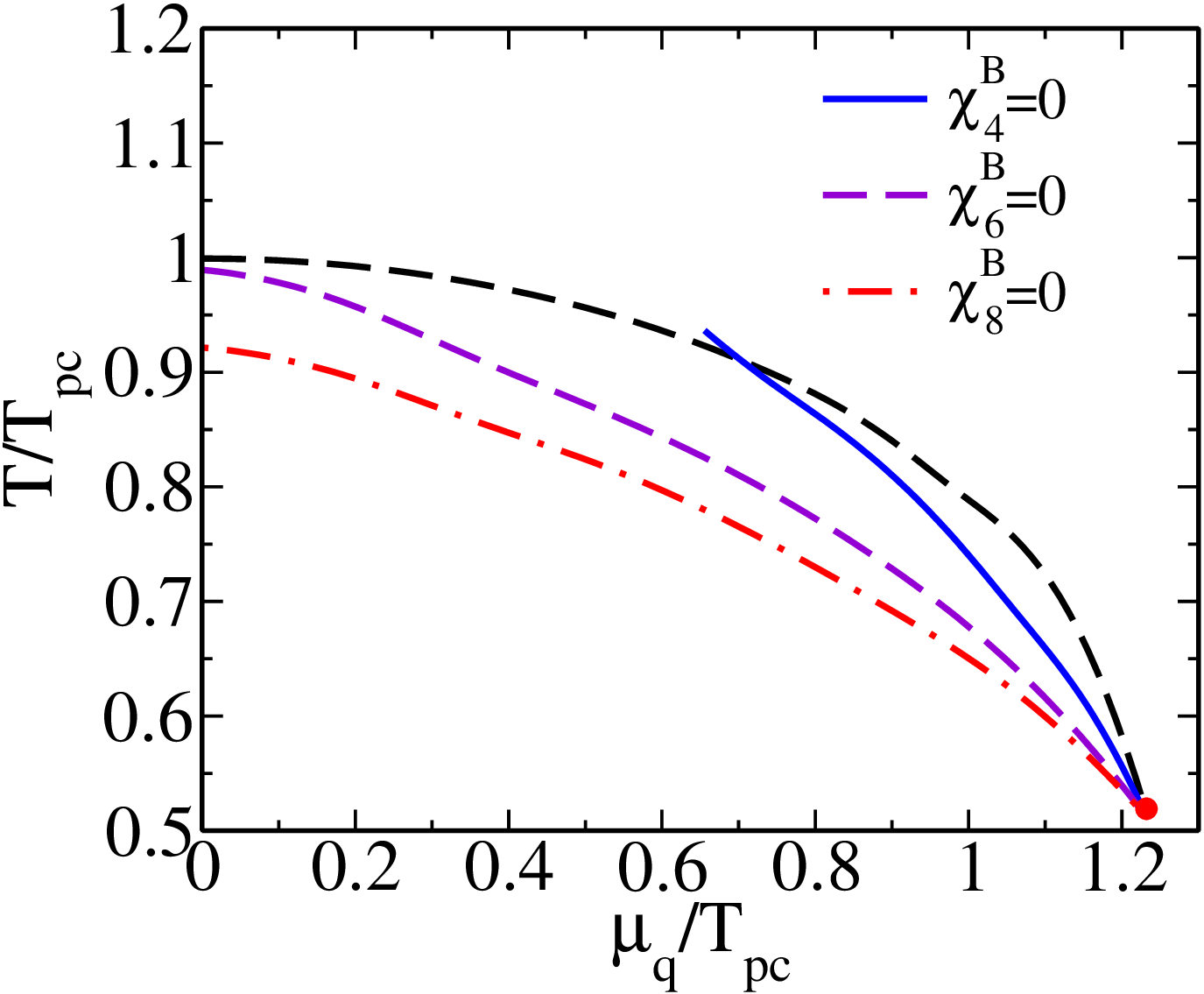}
  \includegraphics*[width=5cm]{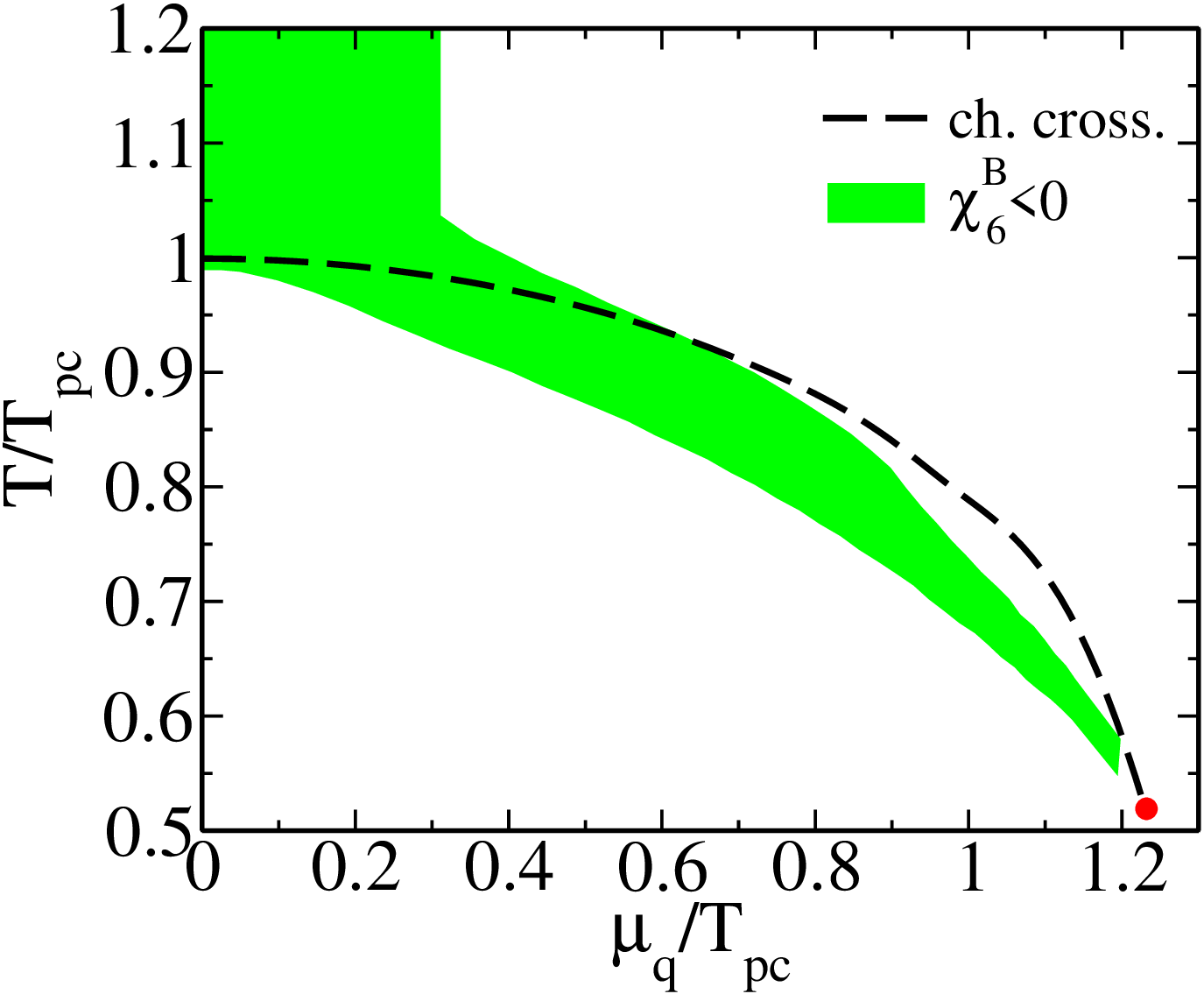}
  \includegraphics*[width=5cm]{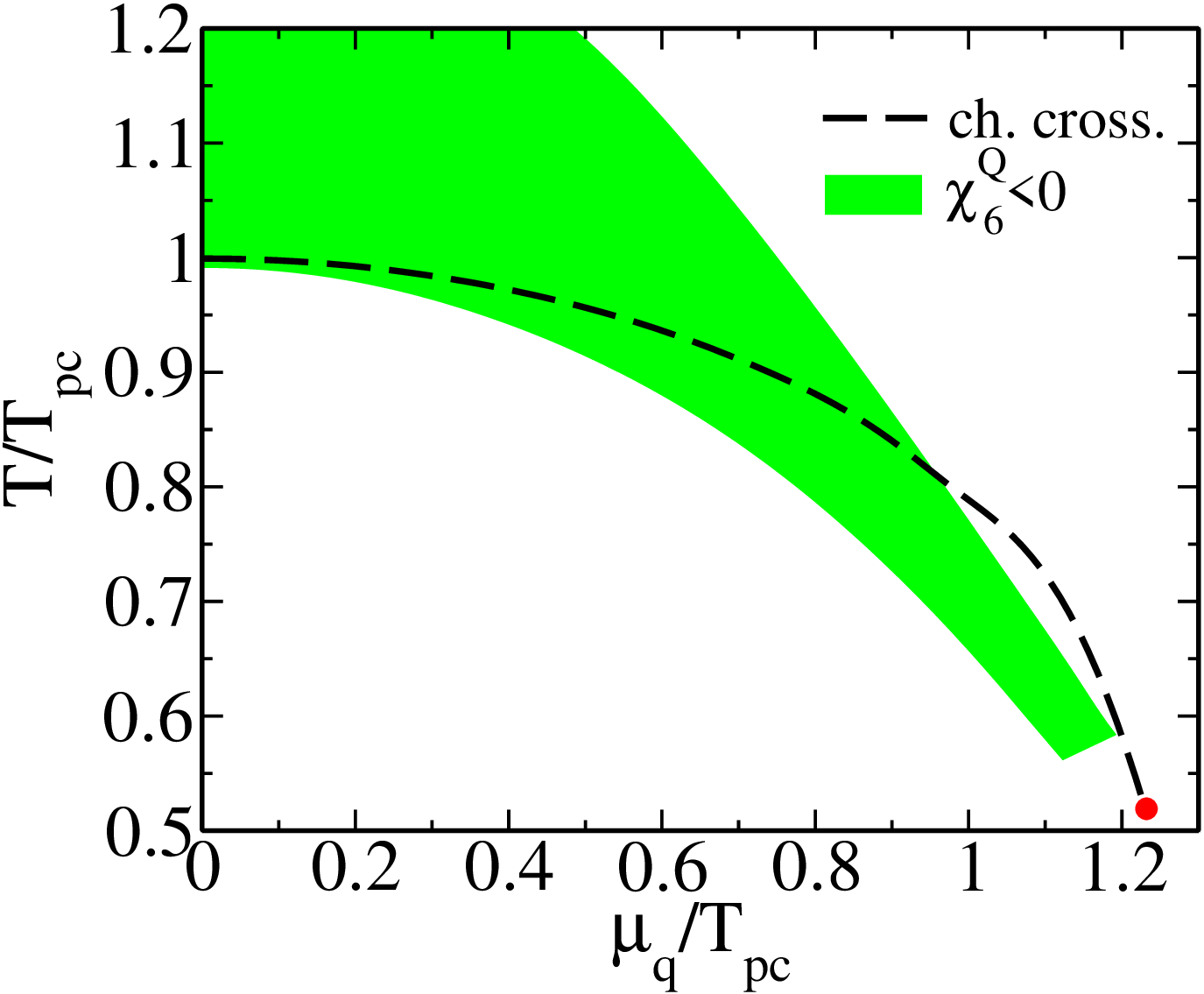}
  \caption { Left: The chiral crossover line (dashed) and the first zeros in $\chi_n^B$ for $n=4,6$ and 8.  Middle: The parameter range for which the sixth order cumulant of net-baryon number fluctuations  is negative. Right:  The parameter range for which the sixth order cumulant of electric charge fluctuations  is negative.  }
  \label{fig:N}
\end{figure*}


\end{document}